\documentclass[sigconf]{acmart}

\renewcommand\footnotetextcopyrightpermission[1]{} % removes footnote with conference information in first column
\usepackage{graphicx}
\usepackage{multirow}
\graphicspath{ {./figures/} }
\usepackage{enumitem}
\usepackage[font=small,skip=0pt]{caption}
\usepackage{makecell}

\usepackage{titlesec}
\usepackage{xspace}

\usepackage{changepage}
\usepackage{subfigure}

\titlespacing\section{0pt}{12pt plus 4pt minus 2pt}{0pt plus 2pt minus 2pt}
\titlespacing\subsection{0pt}{12pt plus 4pt minus 2pt}{0pt plus 2pt minus 2pt}
\titlespacing\subsubsection{0pt}{12pt plus 4pt minus 2pt}{2pt plus 2pt minus 2pt}

\usepackage{collcell}
\usepackage{array}
\usepackage{tikz}
\usepackage{pgfkeys}
\usepackage{graphicx}

\definecolor{aoenglish}{rgb}{0.0, 0.5, 0.0}
\definecolor{celadon}{rgb}{0.67, 0.88, 0.69}

\newcommand*{\MinNumber}{-0.096}%
\newcommand*{\MidNumber}{0.452} %
\newcommand*{\MaxNumber}{1.0}%

\newcommand{\ApplyGradient}[1]{%
        \if\relax\detokenize{#1}\relax% empty cell
        \else
            \ifdim #1 pt > \MidNumber pt
                \pgfmathsetmacro{\PercentColor}{max(min(100.0*(#1 - \MidNumber)/(\MaxNumber-\MidNumber),100.0),0.00)} %
                \hspace{-0.33em}\colorbox{aoenglish!\PercentColor!yellow}{#1}
            \else
                \pgfmathsetmacro{\PercentColor}{max(min(100.0*(\MidNumber - #1)/(\MidNumber-\MinNumber),100.0),0.00)} %
                \hspace{-0.33em}\colorbox{red!\PercentColor!yellow}{#1}
            \fi
        \fi
}

\def\BibTeX{{\rm B\kern-.05em{\sc i\kern-.025em b}\kern-.08emT\kern-.1667em\lower.7ex\hbox{E}\kern-.125emX}}
    
\newcommand{\yashar}[1]{\textcolor{green!55!blue}{{\bf [Yashar: }{#1}{\bf ]}}}

\acmYear{2019}
\setcopyright{none}

\acmConference[RecSys ImpactRS '19, 16 - 20 September, 2019, Copenhagen, Denmark]{The 1st Workshop on the Impact of Recommender Systems}{September, 2019}{Copenhagen, DK}
\acmDOI{}
\acmISBN{978-1-4503-9999-9/18/06}
\newtheorem{definition}{Definition}

\begin{document}

% \title{A Study of the Impact of User-Item Attacks on Collaborative Filtering Systems}

\title{Assessing the Impact of a User-Item Collaborative Attack on Class of Users}
\titlenote{Copyright 2019 for this paper by its authors. Use permitted under Creative Commons License Attribution 4.0 International (CC BY 4.0).\\Presented at the ImpactRS workshop held in conjunction with the 13th ACM Conference on Recommender Systems (RecSys), 2019, in Copenhagen, Denmark.}
%
% The "author" command and its associated commands are used to define the authors and their affiliations.
% Of note is the shared affiliation of the first two authors, and the "authornote" and "authornotemark" commands
% used to denote shared contribution to the research.
% \author{Felice Antonio Merra}
% \email{felice.merra@poliba.it}
% \affiliation{%
%  \institution{Politecnico di Bari}
%   \yashar{Politecnico di Bari is probably better, will tell you why.}}
%  \streetaddress{Via E. Orabona, 4}
%  \city{Bari}
%  \state{Italy}
%  \postcode{70126}
\author{Yashar Deldjoo}
\affiliation{%
  \institution{Polytechnic University of Bari}
  \city{Bari}
  \country{Italy}}
\email{yashar.deldjoo@poliba.it}

\author{Tommaso Di Noia}
\affiliation{%
  \institution{Polytechnic University of Bari}
  \city{Bari}
  \country{Italy}}
\email{tommaso.dinoia@poliba.it}

\author{Felice Antonio Merra}
\affiliation{%
  \institution{Polytechnic University of Bari}
  \city{Bari}
  \country{Italy}}
\email{felice.merra@poliba.it}
\authornote{Authors are listed in alphabetical order. Corresponding author: Felice Antonio Merra (\texttt{felice.merra@poliba.it}).}

% By default, the full list of authors will be used in the page headers. Often, this list is too long, and will overlap
% other information printed in the page headers. This command allows the author to define a more concise list
% of authors' names for this purpose.
\renewcommand{\shortauthors}{ }

\begin{abstract}
 Collaborative Filtering (CF) models lie at the core of most recommendation systems due to their state-of-the-art accuracy. They are commonly adopted in e-commerce and online  services  for their impact on  sales  volume and/or  diversity, and their impact on companies' outcome.
 However, CF models are only as good as the interaction data they work with. As these models rely on outside sources of information, counterfeit data such as user ratings or reviews can be injected by attackers to manipulate the underlying data and alter the impact of resulting recommendations, thus implementing a so-called shilling attack.  
 While previous works have focused on evaluating shilling attack strategies from a global perspective paying particular attention to the effect of the size of attacks and attacker's knowledge, in this work we explore the effectiveness of shilling attacks under novel aspects. First, we investigate the effect of attack strategies crafted on a \textit{target user} in order to push the recommendation of a low-ranking \textit{item} to a higher position, referred to as \textit{user-item attack}. Second, we evaluate the effectiveness of attacks in altering the impact of different CF models by contemplating the \textit{class} of the target user, from the perspective of the richness of her profile (i.e., \textit{slightly-active} v.s. \textit{highly-active} user). Finally, similar to previous work we contemplate the \textit{size of attack} (i.e., the amount of fake profiles injected) in examining their success. %\felice{it is ok}

 %The goal of this work is to investigate the effectiveness of shilling attacks under novel aspects. We focus our attention on the effect of attacks against a \textit{target user} in order to push the recommendation of a low-ranking \textit{item} to an higher position, we called this attack like \textit{user-item attack}.
%Toward this goal, we evaluate the effectiveness of attacks in altering the impact of different baseline CF models by contemplating the \textit{category} of the target-user (i.e., \textit{slightly-active} or \textit{highly-active} user)  the \textit{quantity} (i.e., amount of fake profiles injected).

%works have been focused on assessing the impact of shilling attacks on different recommender models (often CF models) by examining the success of attacks from the perspective of individual users.

The results of experiments on two widely used datasets in business and movie domains, namely Yelp and MovieLens, suggest that highly-active and slightly-active users exhibit contrasting behaviors in datasets with different characteristics.

%~\felice{It is ok. I update your sentence where you use 4 times the adjective different}
%\yashar{different user classes exhibit different behaviors in two datasets with different characteristics}.

% on a single user for a specific item is effectively altered under particular attacks configurations and user property.

% the \textit{quality} (i.e., type of knowledge on the victim user) and .

\end{abstract}

\maketitle

\section{Introduction and Related Work}
% \yashar{I think we can reduce the introduction and include 4 attack types and 4 rec model or 2 attack types and 4 rec models.}
Collaborative filtering (CF) models are a crucial component in many real-world recommendation services due to their state-of-the-art accuracy. Considering their widespread popularity and adoption in the industry, the output of these models can impact many decision qualities in different application scenarios~\cite{shi2014collaborative, DBLP:journals/dsonline/LindenSY03, DBLP:reference/sp/AmatriainB15}. 
The open nature of CF models, which rely on user-specified judgments (e.g., ratings or reviews) to build user profiles and compute recommendation, can be used in the hand of adversaries to manipulate the underlying data and affect the impact of recommendation, a phenomenon commonly referred to as shilling attacks~\cite{mobasher2005effective, DBLP:journals/air/GunesKBP14}. The attacker may manipulate the recommender for positive motivations, like outcomes improvement, or malicious, like reducing the user's loyalty to a competitor.
%\felice{It is ok}
% \yashar{This sentence is very long, I don't think we need to change the introuction that we wrote in RecSys, Please disconnect long sentences and let's try shorter sentence }~\felice{I took it directly from the recsys version}

In this direction, first works~\cite{DBLP:conf/www/LamR04,DBLP:journals/toit/OMahonyHKS04, mobasher2005effective} focused on different profile injection strategies by analyzing and classifying them on the required effort and amount of attacker's knowledge to craft successful attacks. These works have been followed by multiple studies on the evaluation of the robustness~\cite{bhaumik2006securing, DBLP:conf/ictai/ChengH10, DBLP:journals/toit/MobasherBBW07} of different CF models and detection strategies~\cite{DBLP:journals/jiis/YangC17, DBLP:conf/widm/ChiritaNZ05, DBLP:journals/umuai/MehtaN09}. 
The robustness analysis in surveys~\cite{DBLP:journals/expert/MobasherBBS07, DBLP:journals/air/GunesKBP14} shows that Item-\textit{k}NN is more robust than User-\textit{k}NN and model-based CF are generally more resistant to shilling attacks than conventional nearest neighbor-based algorithms.

% \yashar{-\textbf{Please check the following part}- }\\
One common characteristic of the previous literature on shilling attacks on CF-RS is their focus on assessing the global impact of shilling attacks on different CF models by examining the success of attacks from the perspective of attacker's knowledge and the size of attack (i.e. the number of shilling profiles)~\cite{DBLP:journals/air/GunesKBP14}. In the present work instead, we investigate the effectiveness of an attack on a target-item of a target-user, namely \textit{user-item attack}, with a novel point of attention focused on influence of the attack on the \textit{classes of attacked users} in particular \textit{highly-active (HA) user} and \textit{slightly-active (SA) user}. 
%\felice{Ok} 
% \yashar{till here}

%\yashar{ Let's use some connecting sentences: We argue this to be important for targeted attacks on group/classes ... }

%the impact of shilling attacks from the perspective of the global effect of attacks on the recommender model by carefully looking to different attack strategies and the importance of attacker's knowledge and the size of attack (i.e. the number of shilling profiles)~\cite{DBLP:journals/air/GunesKBP14}. 
%These works have been focused on assessing the impact of shilling attacks on different recommender models (often CF models) by examining the success of attacks from the perspective of individual users.

% \yashar{Let us leave the word carefully. What about simplifying this abit by saying: \dquotes{that they have been focused on assessing the impact of shilling attacks on different recommender models (often CF models) by examining the success of attacks from the perspective of individual users.}}.~\felice{It's fine. Ok!}

The application scenario for class-based study of attacks on RS may span in different domains. 
As an example, a restaurant owner may wish to diminish the trust on a target user of a competitor by pushing a low-ranked product \textit{for the specific user}. The same argument can be made for new users. An attacker may be interested in pushing or nuking, particular products with the objective of modifying the impact of a recommender system in order to affect future interactions of the new user.

%\yashar{ RQs can be improved with respect to the final results.}
The leading reserach questions of this work are then:
\begin{itemize}
    \item \textbf{RQ1:} From a global perspective, what is the impact of user-item attack on \textit{classes of users} such as slightly-active and highly-active users?
    \begin{itemize}
        \item Could attacks be tailored to have a higher impact on a particular class of users? 
        \item Which factors play a role on the impact of such an attack?
    \end{itemize}
    \item  \textbf{RQ2:} From a local perspective, how do CF recommendation model work differently under user-item attacks by looking to user-classes?
    % \item  \textbf{RQ2:} From a local perspective, how do different CF recommendation models compare with each other of such attack?
\end{itemize}

% \yashar{We shall be reminded that instead of focusing on individual users, we operationalize the above reach questions by providing a relative comparison of the impact of RS for  }
The remainder of the paper is structured as follows.
Section~\ref{sec:U_I_attack} presents the evaluation protocol and datasets description we used in our experimental evaluation.
Section~\ref{sec:results} reports on the  results and their discussion. Section~\ref{sec:conclusion} concludes the paper and introduces future perspectives.

\section{User-Item attack modeling and evaluation}\label{sec:U_I_attack}
In this section, we discuss our evaluation protocol for a user-item attack modeling and the corresponding evaluation setup.

\subsection{Evaluation Protocol}
\label{sec:Eval_prot}

In order to test the effects of a user-item attack on attacked user classes, an extensive set of experiments has been carried out with respect to three dimensions: (i) the attack strategy (type and quantity of injected profiles), (ii) core CF recommendation model and (iii) the user classes.
The experimental evaluation has been executed on two well-known datasets, MovieLens-1M (ML-1M) and Yelp (described in Section~\ref{sec:data_descr}). 
%In the following, we describe each of these dimensions:
%~\felice{it is ok}
% \yashar{Here we just need to name it, also let us not use new terms - simply say User Classes (corresponding to section 2.1.3)}. 

\subsubsection{Attack Strategies.}\label{subsec:attack}
We have implemented two attack strategies to craft shilling profiles (SP) in order to model different level of attacker's capability. Given a user profile $P(u) = \{r_{i_1}, \ldots ,r_{i_n}\}$ (consisting of a set of items rated by user $u$), we consider the items in $P(u)$ in the form of: \textit{selected items} ($I_{S}$), \textit{filler items} ($I_{F}$), \textit{target item} ($I_{T}$) previously identified in ~\cite{burke2005identifying}, with $|I_{S}|+|I_{F}|+|I_{T}| = |P(u)|$. 
The items in the set $I_{F}$ are selected randomly in order to obstruct detection of an SP while the only element in $I_{T}$ is the item that the attacker wants to push, or nuke. Here we focus on two strategies to build $I_{S}$, which lies at the core of a shilling profile generation. The number of items in a shilling profile is close to the mean value of the number of rating in the dataset. We execute two types of attacks: 
% \textit{user-and-model aware attack} and \textit{user-neighbor aware attack}. 
%\yashar{Check the following attacked carefully - the reviewers asked us to specify the policy of attackers}
\begin{itemize}
    \item \textbf{User-and-Model aware attack}  (\textit{UMA}) assumes a partial knowledge of some victim preferences.
    The attacker creates a new profile, called \textit{seed profile}, on the system with these preferences and uses the recommendation systems to receive recommendations. The recommendations are then used to fill $I_{S}$ with high ratings. This type of attack is inspired by the \textit{probe attack}~\cite{burke2005identifying, DBLP:journals/air/GunesKBP14, Aggarwal01}. In the probe attack, the seed profile is created by the adversary and the recommendations generated by the recommender system are used to learn related items and their ratings in order to built up shilling profiles very similar to existing users in the system.
    These items constitute the 50\% of each shilling profile. 
    \item \textbf{User-Neighbor aware attack} (\textit{UNA})  assumes that the attacker knows some users similar to the victim.
    We employ this attack by evaluating the \textit{k-nearest neighbor} users of each victim\footnote{experiment setting: $k=50$, \textit{similarity metric} = cosine similarity. } and selecting the most rated items in the neighbor in order to fill $I_{S}$. This attack is a modified version of the \textit{bandwagon or popular attack}~\cite{DBLP:conf/aaai/OMahonyHS05}. While the bandwagon attack sets high ratings on the popular items of the system; the proposed attack sets high ratings on the popular items inside the victim's neighborhood in order in order to inject profiles capable to influence more the victim-s recommendations.
\end{itemize}
% \felice{Yashar: could you check this last sentence?}
% Both of these attack strategies assume that the attacker owns a partial knowledge on his, or her, victims,  which is closely linked to the collaborative nature of recommendation process, i.e., UMA exploits collaborative recommendations and UNA uses items liked by user in victim's neighbor.
We executed experiments with different size of injected profiles, which are classified in \textit{small-size} attacks by averaging results of attacks with $2$, $10$, $20$, $50$ shilling profiles and \textit{large-size} attacks by averaging attacks with $200$ and $500$ injected profiles.
%~\felice{It is ok. Update in Averaging!} 
% \yashar{Should we say: the reported results are average values obtained across the above values?}

\subsubsection{CF Models.} 
In our evaluation, we compared the vulnerability/robustness of the following CF models:

\noindent \textbf{User-$k$NN}~\cite{DBLP:conf/uai/BreeseHK98}: user-based $k$-nearest-neighbor ($k$NN) method. In our experiments, we set the number of neighbors $k$ to 20~\cite{mobasher2005effective}. 

\noindent \textbf{Item-$k$NN}~\cite{DBLP:conf/www/SarwarKKR01}: item-based $k$NN method. Also in this case,  the number of neighbors $k$ has been set equal to 20.

\noindent \textbf{BPR-SLIM}~\cite{DBLP:conf/icdm/NingK11}: Sparse LInear Method (SLIM) is an item-item model that models the estimation of unknown user-item rating as a regression problem. It learns a sparse aggregation coefficient matrix from aggregated users' preferences. This model allows the system to capture correlations between items. BPR-SLIM uses the BPR optimization criterion.\footnote{The computation of the CF comparative models has
been done with the publicly available software library MyMediaLite~\url{http://www.mymedialite.net/}. We used default parameters for both BPR-MF and BPR-SLIM.}

\noindent \textbf{BPR-MF}~\cite{DBLP:conf/uai/RendleFGS09}: This method uses matrix factorization (MF) as its underlying core predictor and optimizes it with Bayesian Personalized Ranking (BPR) objective function.

These CF models stand for state-of-the-art models for the item recommendation task, each using a different prediction concept, allowing us to study the impact of different attack strategies from multiple viewpoints.   
\subsubsection{User Classes.}
 Given that CF models only rely on user preference scores (i.e., ratings) to compute recommendation, we hypothesize that it is relevant to investigate the impact of different attack strategies with respect to the victim user's level of activity, i.e. the richness of her profile, calculated on the basis of the number of ratings available in her profile. %In order to have a more intuitive sense about how different attack strategies/configurations presented in Section~\ref{subsec:attack} are effective, we study the success of an attack with respect to users' activeness, where the latter is defined on the number of ratings they provide. 
To this aim, we define two classes of users:
\begin{itemize}
    \item \textbf{Highly-active (HA) users} are defined as users who have a number of ratings greater than the second quartile of the number of ratings for each user in the dataset. 
    \item \textbf{Slightly-active (SA) users} are defined as users who have a number of ratings lower than the second quartile.
\end{itemize}

\subsubsection{Evaluation Metric. }
%  We use a \textit{modified version} of \textit{Hit-Ratio}~\cite{DBLP:conf/sigir/MehtaN08} to measure the fraction of successful user-item attacks which alter the impact of the recommendation engine.
Several metrics have already been proposed to evaluate malicious attacks. For example,~\cite{DBLP:journals/toit/OMahonyHKS04} proposes the prediction shift (PS) which estimates the success of an attack by measuring the prediction difference before and after the attack~\cite{zhang2014addressing}. It has been identified that a strong PS does not necessarily implies an effective attack result~\cite{mobasher2006model}. From the perspective of the attacker, the ideal goal in a push attack is to increase the chance of a desired item being recommended after the attack than before. We use a \textit{modified version} of \textit{Hit-Ratio}~\cite{DBLP:conf/sigir/MehtaN08} to measure the fraction of successful attacks on a set of different user-item pairs.
\begin{definition}
Let $u$ be the user under attack and $i$ be the targeted item that the attacker wants to push/appear in the top-$k$ recommendations of $u$.  
Let $top_u^k$ be the top-$k$ recommendations of $u$. Let $\phi(i, top_u^k)$ be the function to evaluate the effectiveness on an attack on $(u, i)$.
If $i$ is pushed in the top-$k$ then $\phi(i, top_u^k) =1$ (successful attack), otherwise $\phi(i, top_u^k) = 0$ (unsuccessful attack). 
Let $S$ be the set of $(u, i)$ user-item pairs under attack. $HR@k$ is defined as the fraction of successful attacks on each $(u, i) \in S$.
\begin{equation}
    HR@k = \dfrac{\sum_{(u, i) \in S}{}{\phi(i, top_u^k)} }{|S|}
    \label{eq:hr@k}
\end{equation}
where $|S|$ is the number of $(u, i)$ pairs over which $HR@k$ is measured.
\end{definition}

\subsection{Data Descriptions}\label{sec:data_descr}
We conducted experiments on two well-known datasets, MovieLens 1M~\cite{DBLP:journals/tiis/HarperK16} and Yelp~\cite{AdvRec01,DBLP:conf/sigir/HeZKC16}. The datasets represent different item recommendation scenarios for movie and business domains and have data densities which are approximately 40 times different from each other. 
% This allows us to test the vulnerability/robustness of different CF models against the proposed shilling attacks strategies under (datasets with) diverse sparsity levels.\footnote{Note that the term sparsity and density have an equal sense as density = 1-sparsity.}
Table~\ref{tbl:dataset_char} summarizes the statistics of the two datasets (after pre-processing).

 \begin{table}[h]
  \caption{Characteristics of the  dataset used in the offline experiment: $\left| \mathcal{U} \right|$ is the number of users, $\left| \mathcal{I} \right|$ the number of items, $\left| \mathcal{R} \right|$ the number of ratings}
  \label{tbl:dataset_char}
  \begin{tabular}{ccccc}
    \toprule
    \textbf{Dataset} & $\left| \mathcal{U} \right|$ & $\left| \mathcal{I} \right|$ & $\left| \mathcal{R} \right|$  & $\frac{\left|\mathcal{R} \right|}{\left| \mathcal{I} \right| \cdot \left| \mathcal{U} \right|} \times 100$ \\
    \midrule
    \textbf{ML-1M} & 6040 & 3706 & 1000209 & 4.468\%\\
    \textbf{Yelp} & 5135 & 5163 & 24809 & 0.093\%\\
    \bottomrule 
  \end{tabular}
\end{table}

\noindent \textbf{MovieLens-1M}: We used a million-sized version of the dataset ML-1M, which contain 1M ratings of users for items (movies). We used the original ML-1M dataset for the experiments without any filtering.

\noindent \textbf{Yelp}: This dataset contains ratings of users on businesses. We used the pre-processed version of the dataset provided by~\cite{AdvRec01,DBLP:conf/sigir/HeZKC16} with 731K ratings of 25K users for 25K businesses. Given the large size of users and items from which item-item or user-similarities have to be computed, similar to~\cite{DBLP:journals/tkde/AdomaviciusZ15} we extracted a random sample of 5K users and 5K items in order to speed up the experiments. The resulted dataset contains 24.8K ratings with data density (0.110\%), which is comparable with the one before filtering (0.093\%).

\section{Results and Discussion}
\label{sec:results}
%\yashar{I am here xxx THIS SECTION IS LOCKED XXX}
In order to validate the empirical impact of the under study attack types \textit{on different classes of users}, an extensive set of experiments has been carried out with respect to the dimensions introduced in Section  \ref{sec:Eval_prot}. The final results are presented in Table~\ref{tab:results} and discussed from the following viewpoints:

\begin{itemize}
    \item  A \textit{global analysis} of the impact of  attacks on user classes (cf. Section~\ref{subsec:global_impact})
     \item  A \textit{fine-grained analysis} of the impact of  attacks on user classes by looking into the CF models and attack types. (cf. Section~\ref{subsec:local})
    %  \yashar{may improve this one}
\end{itemize}
We present each of these analysis viewpoints in the following subsections.

\begin{table*}[ht]
    \caption{HR@10 for \textit{small-size} and \textit{large-size} attacks with respect to the class of user, \textit{slightly-active} and \textit{highly-active}, and the CF model. The \textit{user-class impact} $r$ is the ratio of $HR_{HA}$ value to $HR_{SA}$. (Abbreviations: HA $\rightarrow$ Highly Active, SA$\rightarrow$Slightly Active)}
    % \yashar{The very lowest line seem to be very bold. In general we can imrpove the visuluation later, let us improve the paper.}}
    % The last row is the standard deviation ($\sigma$) of $r$ for each model under the same size of attack.
    % \begin{tabular}{|l|l|l|l|l|l|l|l|l|l|l|l|l|l|}
    \begin{tabular}{@{\extracolsep{1pt}}lllllllcllllcc@{}}

    \Xhline{3\arrayrulewidth}
     &  & \begin{tabular}[c]{@{}l@{}}\end{tabular} & \multicolumn{5}{c}{\textbf{Small-size attacks}} 
     &  \multicolumn{5}{c}{\textbf{Large-size attacks}} &  \\ 
    %  \cline{4-8} \cline{9-13}
     \cmidrule(lr){4-8} \cmidrule(lr){9-13} 
    \textbf{Dataset} & \begin{tabular}[c]{@{}l@{}}\textbf{CF}/\textbf{Attack}      \end{tabular} & \begin{tabular}[c]{@{}l@{}}              \end{tabular} & \begin{tabular}[c]{@{}l@{}}{\small U-\textit{k}NN}           \end{tabular} & \begin{tabular}[c]{@{}l@{}}{\small I-\textit{k}NN}        \end{tabular} & \begin{tabular}[c]{@{}l@{}}{\small BPR}\\ {\small SLIM}           \end{tabular} & \begin{tabular}[c]{@{}l@{}}{\small BPR} \\{ \small MF}           \end{tabular}
    & \textbf{mean} & 
    \begin{tabular}[c]{@{}l@{}}{\small U-\textit{k}NN}      \end{tabular} & \begin{tabular}[c]{@{}l@{}}{\small I-\textit{k}NN}               \end{tabular} & 
    \begin{tabular}[c]{@{}l@{}}{\small BPR}\\ {\small SLIM}           \end{tabular} & 
    \begin{tabular}[c]{@{}l@{}}{\small BPR} \\{ \small MF}                          \end{tabular} 
    & \textbf{mean} &
    \begin{tabular}[c]{@{}c@{}}\textbf{overall}\\ \textbf{mean}\end{tabular} \\ \hline
    
    \multirow{7}{*}{\textbf{Yelp}} & \multirow{3}{*}{\textbf{UMA}} & SA  & 0.750 & 0.067 & 0.225 & 0.108 & \textbf{0.288} & 0.967 & 0.184 & 0.500 & 0.533 & \textbf{0.546} & \textbf{0.417} \\ 
     &  & HA  & 0.800 & 0.350 & 0.492 & 0.117 & \textbf{0.440} & 1.000 & 0.667 & 0.784 & 0.584 & \textbf{0.758} & \textbf{0.599} \\ 
     &  & $r$ & \textit{1.067} & \textit{5.243} & \textit{2.184} &\textit{ 1.079} & \textbf{2.393} &\textit{ 1.034} & \textit{3.632} &\textit{ 1.567} & \textit{1.095} & \textbf{1.832} & \textbf{2.113} \\ \cline{2-14} 

     & \multirow{3}{*}{\textbf{UNA}} & SA  & 0.850 & 0.625 & 0.792 & 0.400 & \textbf{0.667} & 1.000 & 0.834 & 1.000 & 1.000 & \textbf{0.958} & \textbf{0.813} \\ 
     &  & HA  & 0.875 & 0.742 & 0.850 & 0.433 & \textbf{0.725} & 1.000 & 0.850 & 1.000 & 1.000 & \textbf{0.963} & \textbf{0.844} \\ 
     &  & $r$ & \textit{1.029 }& \textit{1.186} & \textit{1.074} & \textit{1.082} & \textbf{1.093} & \textit{1.000} & \textit{1.020} & \textit{1.000} & \textit{1.000} & \textbf{1.005} & \textbf{1.049} \\ 
     
    \Xhline{2\arrayrulewidth}

    \multirow{7}{*}{\textbf{ML-1M}} & \multirow{3}{*}{\textbf{UMA}} & SA  & 0.302 & 0.155 & 0.267 & 0.121 & \textbf{0.211} & 0.897 & 0.086 & 0.586 & 0.328 & \textbf{0.474} & \textbf{0.343} \\ 
     &  & HA  & 0.092 & 0.108 & 0.159 & 0.125 & \textbf{0.121} & 0.383 & 0.150 & 0.350 & 0.284 & \textbf{0.292} & \textbf{0.206} \\ 
     &  & $r$ &\textit{ 0.303} & \textit{0.698} & \textit{0.593} & \textit{1.037} & \textbf{0.658} & \textit{0.427} & \textit{1.744} & \textit{0.597} & \textit{0.866} & \textbf{0.909} & \textbf{0.783} \\ \cline{2-14} 
     & \multirow{3}{*}{\textbf{UNA}} & SA  & 0.621 & 0.302 & 0.595 & 0.164 & \textbf{0.420} & 1.000 & 0.897 & 1.000 & 0.811 & \textbf{0.927} & \textbf{0.673} \\ 
     &  & HA  & 0.459 & 0.133 & 0.250 & 0.183 & \textbf{0.256} & 1.000 & 0.800 & 0.800 & 0.600 & \textbf{0.800} & \textbf{0.528} \\
     
     &  & $r$ & \textit{0.739} & \textit{0.442} & \textit{0.421} & \textit{1.121} & \textbf{0.680} & \textit{1.000 }& \textit{0.892} & \textit{0.800} & \textit{0.740} & \textbf{0.858} & \textbf{0.769} \\ 
     
     \Xhline{2\arrayrulewidth}
     
    %  &  & \textit{$\sigma(r)$ } & 0.353 & 2.255 & 0.794 & 0.034 & & 0.293 & 1.264 & 0.418 & 0.155 &  &   \\ 
     
    % \Xhline{3\arrayrulewidth}

    \end{tabular}
    \label{tab:results}
\end{table*}

\subsection{Global impact of attacks on user classes}\label{subsec:global_impact}
%\yashar{This section is almost stable, may need a quick check}\\
The goal of this analysis is to answer the first research question related to the global assessment on the effectiveness of user-item attack with respect to the identified users classes. We use the term global here, since in this analysis we would like to free our attention from the impact of attacks on CF models, attack quality (type) and/or quantity as they have been largely addressed in previous works~\cite{DBLP:journals/expert/MobasherBBS07, DBLP:journals/toit/MobasherBBW07, DBLP:journals/air/GunesKBP14}. Instead, we examine the impact of attacks on the dimension of user classes by looking into the aggregate mean values computed across CF models on the two datasets we adopted in our experimental evaluation.

A general observation for the results in Table~\ref{tab:results} is that larger-size attacks reach higher level of effectiveness on both classes of users (highly-active and slightly-active) in comparison with smaller-size attacks. For example, on the Yelp dataset, the average HR@10 for UNA attack on highly-active users (across CF models) is 0.256 for small-size attacks, while it is 0.800 for large-size attacks, a difference of approximately three times. 
The same pattern of results is obtained in other experimental cases. These results are in line with those presented in previous works~\cite{DBLP:journals/toit/MobasherBBW07, DBLP:journals/expert/MobasherBBS07}. 

Our objective here is to study the impact of different attack strategies on user classes. For this purpose, we define the variable $r = \frac{HR_{HA}}{HR_{SA}}$ and refer to it as \textit{user-class attack impact} ---i.e., the impact of an attack on highly-active users in comparison with slightly-active users. Different values for $r$ are interpreted as in the following:

%fixed an attack type and a CFodel, let $\overline{HR}$ the average hit ratio, then \yashar{Felice, the latter is not a sentence, please when you change sth leave a comment, we may not be able to come back to them again.}~\felice{How can we define r?}

%The impact of attack on highly-active users in comparison with slightly-active users is interpreted as following:
%$r= 0.800 \textbackslash 0.750 = 1.067$ for UMA attack executed on User-\textit{k}NN model in the case of Yelp dataset.

%Fixed the model and a type of attack, we define the variable $r$ as the ratio between the average hit ratio between large-size attack and small-size attack, e.g., $r= 0.800 \textbackslash 0.750 = 1.067$ for UMA attack executed on User-\textit{k}NN model in the case of Yelp dataset.

\begin{itemize}
    \item $r = 1$: the attack has an \textit{equal impact} on highly-active and slightly-active users.
    \item  $r > 1$: the attack has an \textit{unequal impact} on highly-active w.r.t slightly-active users. The impact of attack on \textit{highly-active users} is \textit{relatively} higher in comparison with slightly-active users.\footnote{This is equal to say, slightly-active users are relatively more immune to the attack w.r.t. highly-active users.}
     \item $r < 1$: the attack has \textit{an unequal impact} on highly-active v.s. slightly-active users. The impact of attack on \textit{slightly-active users} is \textit{relatively} higher in comparison with highly-active users.
\end{itemize}
It is obvious that the larger $r$ deviates from the center point 1, the larger is the attack success in differentiating highly-active with respect to slightly-active users in one of the above-mentioned directions ($r<1$ or $r>1$).
% \felice{I move this part here}
Before starting a deeper analysis of the results we highlight that the most interesting values are in the left portion of Table~\ref{tab:results} (small-size attacks), because when the size of attack is larger the attack reaches the maximum effectiveness, $HR = 1$, independently of user classes. 
% \yashar{we will find a place for this sentence later}~\felice{What do you think if we put it like a footnote of the Table 2?} \yashar{we can remind in the table, but I would say let's also mention in the text.}

By looking at the results for each attack size in Table~\ref{tab:results}, we can see that
the average user-class impact $\bar{r}$ has a value higher than 1 for the Yelp dataset ($\bar{r}>1$),  while a value lower than 1 for the ML-1M dataset ($\bar{r}<1$). These results show that \textit{both attack types have an unequal impact on slightly-active vs highly-active users} as $r \neq 1$. However, \textit{the class of users they have a larger impact on remains largely different and contrasting in the two datasets}.

As an example, in Yelp and for UMA, one can note that for small-size attack $\bar{r} = 2.393$ and for large-size attack $\bar{r} = 1.832$, while the corresponding values on ML-1M are $\bar{r} = 0.658$ and $\bar{r} = 0.909$, respectively. This means that the impact of attacks on user classes is higher on highly-active users on the Yelp dataset ($\bar{r}$>1), differently from ML-1M ($\bar{r}$<1). 
% \yashar{I suggest to use the term ML-1M instead of ML-1M throughout the paper everywhere.}

%\yashar{The following I have improved - but need attention}\\
We conjecture that the above contrasting behaviors are directly linked with the characteristics of the datasets such as their sparsity. As shown in Table~\ref{tbl:dataset_char}, Yelp dataset is approximately 40 times sparser than ML-1M and we consider this difference as the main/possible cause of the contrasting outcomes in tested datasets. We try to provide a possible explanation here. In the more sparse dataset (i.e., the Yelp dataset), users with a small number of ratings (\textit{slightly-active users}) are more immune to attacks because they have a smaller support size of the user profile (i.e., the user profile is not rich enough for the attacker to be able to mimic it in a crafted way). In contrast, highly-active users are more immune to attack in ML-1M with higher density, because their recommendations rely on neighbors with (very) rich user profiles. Put it simply, the crafted attacks need to use a large number of profiles to be able to alter recommendation for the target user. 

%have overwhelmed neighbors that protect users \yashar{Not very clear, could we slightly improve?}.
The insight on sparsity is an important indication that data characteristics are playing a role in the effectiveness of attacks and it motivates further research in this direction.

\subsection{Fine-grained analysis of the impact of attacks on user classes}\label{subsec:local}

The goal of this analysis is to study how different CF models behave against the attacks: which ones have similar performance and which ones have a different performance. 
This study resembles previous work on shilling attacks on CF models. However, we take into account the impact of attack on user classes in this study as well.

% \begin{figure}[t]
%   \includegraphics[width=\linewidth]{figures/Heatmap.png}
%   \caption{Heat-map of Correlation Coefficient ($\rho$) of different measures between CF models for \textit{small-size attacks}: (a) HR@10 on slightly-active Users, (b) HR@10 on Highly-active Users.}
%   \label{fig:correlation}
% \end{figure}

\begin{figure}[t]
  \centering
  \subfigure[HR@10 Slightly-active Users]{%
    \includegraphics[width=0.23\textwidth]{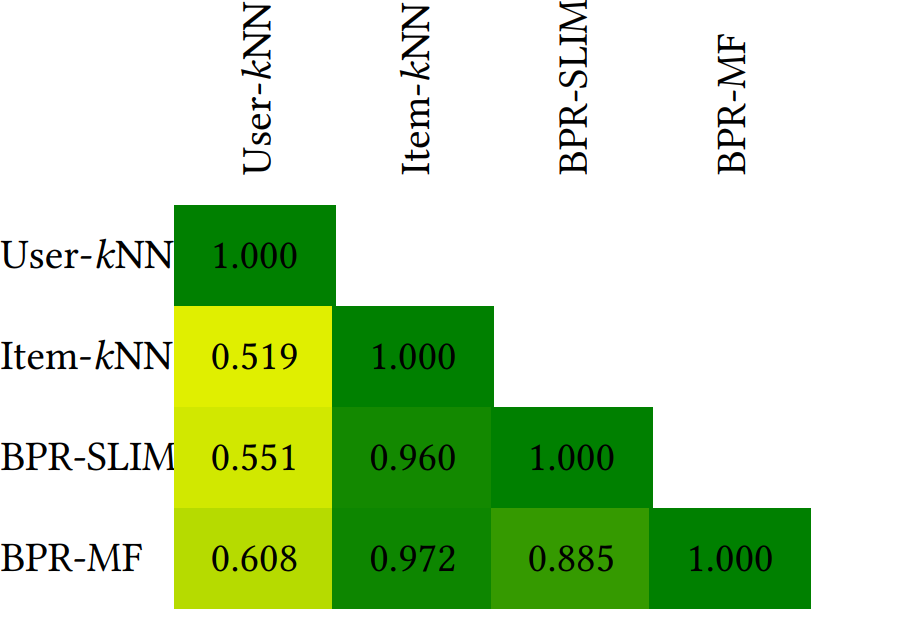}%
    }
    % \hspace{0.2cm}%or more
    \subfigure[HR@10 Highly-active Users]{%
    \includegraphics[width=0.20\textwidth]{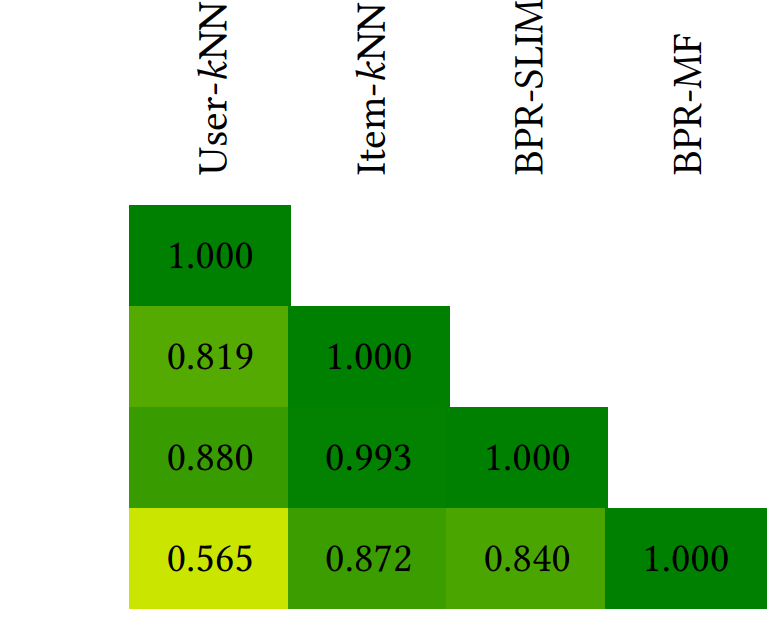}%
  }%  
  \caption{Heat-map of Correlation Coefficient ($\rho$) of different measures between CF models for \textit{small-size attacks}: (a) HR@10 on Slightly-active Users, (b) HR@10 on Highly-active Users.}
\label{fig:correlation}
\end{figure}

Instead of individual CF models performances and attack types, we compute the pairwise Pearson correlation between each pair of analyzed CF models.
% \yashar{explain which ones are taken into account here}~\felice{I consider the Correlation between each pair of models}. 
% \yashar{are we sure this is a table?} 
Figure~\ref{fig:correlation} indicates a strong correlation on HR@10 between BPR-SLIM and Item-\textit{k}NN ($\rho = 0.960$ in Figure~\ref{fig:correlation}a and $\rho = 0.993$ in Figure~\ref{fig:correlation}b). We justify this value by the fact that both CF models exploit the item-item similarity computation. Looking at the correlation values for User-\textit{k}NN in Figure~\ref{fig:correlation}, one can observe a slightly lower correlation in the case of \textit{slightly-active-users} with respect to other models. We think that this phenomenon comes from the fact that tested attack are based on user preferences which gain good effect also with small-size attacks. For instance, HR@10 for Yelp on slightly-active users (0.750 and 0.850) is higher than the mean values with other models for both attack (mean = 0.288 and 0.440). 
We can also observe an interesting behavior when we compare $\rho$ of BPR-MF with BPR-SLIM and Item-\textit{k}NN. Figure~\ref{fig:correlation} (a) and (b) show that HR@10 on both classes of attacked users is highly correlated ($\rho \geq{0.840}$). 
Finally, results in Table~\ref{tab:results} show that BPR-MF is the model that is less influenced by user-classes because the \textit{user-impact factor} is close to 1 for each class of users and attacks. %\yashar{I would ask to read the last paragraph well}

\section{Conclusion and Future Work}\label{sec:conclusion}
This work investigates the effect of user-item attacks on classes of users.
Particularly, we investigated the effectiveness of attacks from a global and local perspective by varying the quality and quantity of attacks, the target user class and the collaborative filtering recommendation model.

Experimental results on Yelp and MovieLens datasets indicate that for Yelp dataset slightly-active users are more immune to shilling attacks than highly-active users, a characteristic that is in contrast with the results on MovieLens dataset where highly-active users are more immune than slightly-active users. 
As datasets have a very different sparsity (Yelp is approximately 40 time more sparse than MovieLens) we will move our future works in analyzing the effectiveness of dataset properties under different attack scenarios. From a local perspective, we evidence that BPR-MF is less influenced than other models when varying user-class and attack types. On the other hand, BPR-SLIM and Item-\textit{k}NN have shown similar behavior related to the effect of attacks on user classes.
In future, we also plan to extend our study by considering more datasets from different domains, exploring in an extensive way the influence of dataset properties, such as sparsity, user and item skewness, rating variance, on the effectiveness of different type of attacks. Also, it is of our interest to consider the impact of various shilling attack types on CF models using item content as side information~\cite{DBLP:journals/umuai/DeldjooDCECSIC19,DBLP:conf/iir/DeldjooSCP18}. 

These studies give important insights on the impact of shilling attacks on recommender systems and provide clues on how to reduce their effectiveness by working on datasets characteristics.

\bibliographystyle{ACM-Reference-Format}
\bibliography{main}


\begin{thebibliography}{30}

%%% the Web address.

\ifx \showCODEN    \undefined \def \showCODEN     #1{\unskip}     \fi
\ifx \showDOI      \undefined \def \showDOI       #1{#1}\fi
\ifx \showISBNx    \undefined \def \showISBNx     #1{\unskip}     \fi
\ifx \showISBNxiii \undefined \def \showISBNxiii  #1{\unskip}     \fi
\ifx \showISSN     \undefined \def \showISSN      #1{\unskip}     \fi
\ifx \showLCCN     \undefined \def \showLCCN      #1{\unskip}     \fi
\ifx \shownote     \undefined \def \shownote      #1{#1}          \fi
\ifx \showarticletitle \undefined \def \showarticletitle #1{#1}   \fi
\ifx \showURL      \undefined \def \showURL       {\relax}        \fi
% invisible to TeX
\providecommand\bibfield[2]{#2}
\providecommand\bibinfo[2]{#2}
\providecommand\natexlab[1]{#1}
\providecommand\showeprint[2][]{arXiv:#2}

\bibitem[\protect\citeauthoryear{Adomavicius and Zhang}{Adomavicius and
  Zhang}{2015}]%
        {DBLP:journals/tkde/AdomaviciusZ15}
\bibfield{author}{\bibinfo{person}{Gediminas Adomavicius} {and}
  \bibinfo{person}{Jingjing Zhang}.} \bibinfo{year}{2015}\natexlab{}.
\newblock \showarticletitle{Improving Stability of Recommender Systems: {A}
  Meta-Algorithmic Approach}.
\newblock \bibinfo{journal}{\emph{{IEEE} Trans. Knowl. Data Eng.}}
  \bibinfo{volume}{27}, \bibinfo{number}{6} (\bibinfo{year}{2015}),
  \bibinfo{pages}{1573--1587}.
\newblock
\urldef\tempurl%
\url{https://doi.org/10.1109/TKDE.2014.2384502}
\showDOI{\tempurl}


\bibitem[\protect\citeauthoryear{Aggarwal}{Aggarwal}{2016}]%
        {Aggarwal01}
\bibfield{author}{\bibinfo{person}{Charu~C. Aggarwal}.}
  \bibinfo{year}{2016}\natexlab{}.
\newblock \bibinfo{booktitle}{\emph{Recommender Systems - The Textbook}}.
\newblock \bibinfo{publisher}{Springer}.
\newblock
\showISBNx{978-3-319-29657-9}
\urldef\tempurl%
\url{https://doi.org/10.1007/978-3-319-29659-3}
\showDOI{\tempurl}


\bibitem[\protect\citeauthoryear{Amatriain and Basilico}{Amatriain and
  Basilico}{2015}]%
        {DBLP:reference/sp/AmatriainB15}
\bibfield{author}{\bibinfo{person}{Xavier Amatriain} {and}
  \bibinfo{person}{Justin Basilico}.} \bibinfo{year}{2015}\natexlab{}.
\newblock \showarticletitle{Recommender Systems in Industry: {A} Netflix Case
  Study}.
\newblock In \bibinfo{booktitle}{\emph{Recommender Systems Handbook}},
  \bibfield{editor}{\bibinfo{person}{Francesco Ricci}, \bibinfo{person}{Lior
  Rokach}, {and} \bibinfo{person}{Bracha Shapira}} (Eds.).
  \bibinfo{publisher}{Springer}, \bibinfo{pages}{385--419}.
\newblock
\showISBNx{978-1-4899-7636-9}
\urldef\tempurl%
\url{https://doi.org/10.1007/978-1-4899-7637-6\_11}
\showDOI{\tempurl}


\bibitem[\protect\citeauthoryear{Bhaumik, Williams, Mobasher, and
  Burke}{Bhaumik et~al\mbox{.}}{2006}]%
        {bhaumik2006securing}
\bibfield{author}{\bibinfo{person}{Runa Bhaumik}, \bibinfo{person}{Chad
  Williams}, \bibinfo{person}{Bamshad Mobasher}, {and} \bibinfo{person}{Robin
  Burke}.} \bibinfo{year}{2006}\natexlab{}.
\newblock \showarticletitle{Securing collaborative filtering against malicious
  attacks through anomaly detection}. In \bibinfo{booktitle}{\emph{Proceedings
  of the 4th Workshop on Intelligent Techniques for Web Personalization
  (ITWP'06), Boston}}, Vol.~\bibinfo{volume}{6}. \bibinfo{pages}{10}.
\newblock


\bibitem[\protect\citeauthoryear{Breese, Heckerman, and Kadie}{Breese
  et~al\mbox{.}}{1998}]%
        {DBLP:conf/uai/BreeseHK98}
\bibfield{author}{\bibinfo{person}{John~S. Breese}, \bibinfo{person}{David
  Heckerman}, {and} \bibinfo{person}{Carl~Myers Kadie}.}
  \bibinfo{year}{1998}\natexlab{}.
\newblock \showarticletitle{Empirical Analysis of Predictive Algorithms for
  Collaborative Filtering}. In \bibinfo{booktitle}{\emph{{UAI} '98: Proceedings
  of the Fourteenth Conference on Uncertainty in Artificial Intelligence,
  University of Wisconsin Business School, Madison, Wisconsin, USA, July 24-26,
  1998}}, \bibfield{editor}{\bibinfo{person}{Gregory~F. Cooper} {and}
  \bibinfo{person}{Seraf{\'{\i}}n Moral}} (Eds.). \bibinfo{publisher}{Morgan
  Kaufmann}, \bibinfo{pages}{43--52}.
\newblock
\showISBNx{1-55860-555-X}
\urldef\tempurl%
\url{https://dslpitt.org/uai/displayArticleDetails.jsp?mmnu=1\&smnu=2\&article\_id=231\&proceeding\_id=14}
\showURL{%
\tempurl}


\bibitem[\protect\citeauthoryear{Burke, Mobasher, Zabicki, and Bhaumik}{Burke
  et~al\mbox{.}}{2005}]%
        {burke2005identifying}
\bibfield{author}{\bibinfo{person}{Robin Burke}, \bibinfo{person}{Bamshad
  Mobasher}, \bibinfo{person}{Roman Zabicki}, {and} \bibinfo{person}{Runa
  Bhaumik}.} \bibinfo{year}{2005}\natexlab{}.
\newblock \showarticletitle{Identifying attack models for secure
  recommendation}.
\newblock \bibinfo{journal}{\emph{Beyond Personalization}}
  \bibinfo{volume}{2005} (\bibinfo{year}{2005}).
\newblock


\bibitem[\protect\citeauthoryear{Cheng and Hurley}{Cheng and Hurley}{2010}]%
        {DBLP:conf/ictai/ChengH10}
\bibfield{author}{\bibinfo{person}{Zunping Cheng} {and} \bibinfo{person}{Neil
  Hurley}.} \bibinfo{year}{2010}\natexlab{}.
\newblock \showarticletitle{Robust Collaborative Recommendation by Least
  Trimmed Squares Matrix Factorization}. In \bibinfo{booktitle}{\emph{22nd
  {IEEE} International Conference on Tools with Artificial Intelligence,
  {ICTAI} 2010, Arras, France, 27-29 October 2010 - Volume 2}}.
  \bibinfo{publisher}{{IEEE} Computer Society}, \bibinfo{pages}{105--112}.
\newblock
\urldef\tempurl%
\url{https://doi.org/10.1109/ICTAI.2010.90}
\showDOI{\tempurl}


\bibitem[\protect\citeauthoryear{Chirita, Nejdl, and Zamfir}{Chirita
  et~al\mbox{.}}{2005}]%
        {DBLP:conf/widm/ChiritaNZ05}
\bibfield{author}{\bibinfo{person}{Paul{-}Alexandru Chirita},
  \bibinfo{person}{Wolfgang Nejdl}, {and} \bibinfo{person}{Cristian Zamfir}.}
  \bibinfo{year}{2005}\natexlab{}.
\newblock \showarticletitle{Preventing shilling attacks in online recommender
  systems}. In \bibinfo{booktitle}{\emph{Seventh {ACM} International Workshop
  on Web Information and Data Management {(WIDM} 2005), Bremen, Germany,
  November 4, 2005}}, \bibfield{editor}{\bibinfo{person}{Angela Bonifati} {and}
  \bibinfo{person}{Dongwon Lee}} (Eds.). \bibinfo{publisher}{{ACM}},
  \bibinfo{pages}{67--74}.
\newblock
\showISBNx{1-59593-194-5}
\urldef\tempurl%
\url{https://doi.org/10.1145/1097047.1097061}
\showDOI{\tempurl}


\bibitem[\protect\citeauthoryear{Deldjoo, Dacrema, Constantin, Eghbal{-}zadeh,
  Cereda, Schedl, Ionescu, and Cremonesi}{Deldjoo et~al\mbox{.}}{2019}]%
        {DBLP:journals/umuai/DeldjooDCECSIC19}
\bibfield{author}{\bibinfo{person}{Yashar Deldjoo},
  \bibinfo{person}{Maurizio~Ferrari Dacrema}, \bibinfo{person}{Mihai~Gabriel
  Constantin}, \bibinfo{person}{Hamid Eghbal{-}zadeh}, \bibinfo{person}{Stefano
  Cereda}, \bibinfo{person}{Markus Schedl}, \bibinfo{person}{Bogdan Ionescu},
  {and} \bibinfo{person}{Paolo Cremonesi}.} \bibinfo{year}{2019}\natexlab{}.
\newblock \showarticletitle{Movie genome: alleviating new item cold start in
  movie recommendation}.
\newblock \bibinfo{journal}{\emph{User Model. User-Adapt. Interact.}}
  \bibinfo{volume}{29}, \bibinfo{number}{2} (\bibinfo{year}{2019}),
  \bibinfo{pages}{291--343}.
\newblock
\urldef\tempurl%
\url{https://doi.org/10.1007/s11257-019-09221-y}
\showDOI{\tempurl}


\bibitem[\protect\citeauthoryear{Deldjoo, Schedl, Cremonesi, and Pasi}{Deldjoo
  et~al\mbox{.}}{2018}]%
        {DBLP:conf/iir/DeldjooSCP18}
\bibfield{author}{\bibinfo{person}{Yashar Deldjoo}, \bibinfo{person}{Markus
  Schedl}, \bibinfo{person}{Paolo Cremonesi}, {and} \bibinfo{person}{Gabriella
  Pasi}.} \bibinfo{year}{2018}\natexlab{}.
\newblock \showarticletitle{Content-Based Multimedia Recommendation Systems:
  Definition and Application Domains}. In \bibinfo{booktitle}{\emph{Proceedings
  of the 9th Italian Information Retrieval Workshop, Rome, Italy, May, 28-30,
  2018.}} \emph{(\bibinfo{series}{{CEUR} Workshop Proceedings})},
  \bibfield{editor}{\bibinfo{person}{Nicola Tonellotto}, \bibinfo{person}{Luca
  Becchetti}, {and} \bibinfo{person}{Marko Tkalcic}} (Eds.),
  Vol.~\bibinfo{volume}{2140}. \bibinfo{publisher}{CEUR-WS.org}.
\newblock
\urldef\tempurl%
\url{http://ceur-ws.org/Vol-2140/paper15.pdf}
\showURL{%
\tempurl}


\bibitem[\protect\citeauthoryear{Gunes, Kaleli, Bilge, and Polat}{Gunes
  et~al\mbox{.}}{2014}]%
        {DBLP:journals/air/GunesKBP14}
\bibfield{author}{\bibinfo{person}{Ihsan Gunes}, \bibinfo{person}{Cihan
  Kaleli}, \bibinfo{person}{Alper Bilge}, {and} \bibinfo{person}{Huseyin
  Polat}.} \bibinfo{year}{2014}\natexlab{}.
\newblock \showarticletitle{Shilling attacks against recommender systems: a
  comprehensive survey}.
\newblock \bibinfo{journal}{\emph{Artif. Intell. Rev.}} \bibinfo{volume}{42},
  \bibinfo{number}{4} (\bibinfo{year}{2014}), \bibinfo{pages}{767--799}.
\newblock
\urldef\tempurl%
\url{https://doi.org/10.1007/s10462-012-9364-9}
\showDOI{\tempurl}


\bibitem[\protect\citeauthoryear{Harper and Konstan}{Harper and
  Konstan}{2016}]%
        {DBLP:journals/tiis/HarperK16}
\bibfield{author}{\bibinfo{person}{F.~Maxwell Harper} {and}
  \bibinfo{person}{Joseph~A. Konstan}.} \bibinfo{year}{2016}\natexlab{}.
\newblock \showarticletitle{The MovieLens Datasets: History and Context}.
\newblock \bibinfo{journal}{\emph{TiiS}} \bibinfo{volume}{5},
  \bibinfo{number}{4} (\bibinfo{year}{2016}), \bibinfo{pages}{19:1--19:19}.
\newblock
\urldef\tempurl%
\url{https://doi.org/10.1145/2827872}
\showDOI{\tempurl}


\bibitem[\protect\citeauthoryear{He, He, Du, and Chua}{He
  et~al\mbox{.}}{2018}]%
        {AdvRec01}
\bibfield{author}{\bibinfo{person}{Xiangnan He}, \bibinfo{person}{Zhankui He},
  \bibinfo{person}{Xiaoyu Du}, {and} \bibinfo{person}{Tat{-}Seng Chua}.}
  \bibinfo{year}{2018}\natexlab{}.
\newblock \showarticletitle{Adversarial Personalized Ranking for
  Recommendation}. In \bibinfo{booktitle}{\emph{The 41st International {ACM}
  {SIGIR} Conference on Research {\&} Development in Information Retrieval,
  {SIGIR} 2018, Ann Arbor, MI, USA, July 08-12, 2018}},
  \bibfield{editor}{\bibinfo{person}{Kevyn Collins{-}Thompson},
  \bibinfo{person}{Qiaozhu Mei}, \bibinfo{person}{Brian~D. Davison},
  \bibinfo{person}{Yiqun Liu}, {and} \bibinfo{person}{Emine Yilmaz}} (Eds.).
  \bibinfo{publisher}{{ACM}}, \bibinfo{pages}{355--364}.
\newblock
\urldef\tempurl%
\url{https://doi.org/10.1145/3209978.3209981}
\showDOI{\tempurl}


\bibitem[\protect\citeauthoryear{He, Zhang, Kan, and Chua}{He
  et~al\mbox{.}}{2016}]%
        {DBLP:conf/sigir/HeZKC16}
\bibfield{author}{\bibinfo{person}{Xiangnan He}, \bibinfo{person}{Hanwang
  Zhang}, \bibinfo{person}{Min{-}Yen Kan}, {and} \bibinfo{person}{Tat{-}Seng
  Chua}.} \bibinfo{year}{2016}\natexlab{}.
\newblock \showarticletitle{Fast Matrix Factorization for Online Recommendation
  with Implicit Feedback}. In \bibinfo{booktitle}{\emph{Proceedings of the 39th
  International {ACM} {SIGIR} conference on Research and Development in
  Information Retrieval, {SIGIR} 2016, Pisa, Italy, July 17-21, 2016}},
  \bibfield{editor}{\bibinfo{person}{Raffaele Perego},
  \bibinfo{person}{Fabrizio Sebastiani}, \bibinfo{person}{Javed~A. Aslam},
  \bibinfo{person}{Ian Ruthven}, {and} \bibinfo{person}{Justin Zobel}} (Eds.).
  \bibinfo{publisher}{{ACM}}, \bibinfo{pages}{549--558}.
\newblock
\showISBNx{978-1-4503-4069-4}
\urldef\tempurl%
\url{https://doi.org/10.1145/2911451.2911489}
\showDOI{\tempurl}


\bibitem[\protect\citeauthoryear{Lam and Riedl}{Lam and Riedl}{2004}]%
        {DBLP:conf/www/LamR04}
\bibfield{author}{\bibinfo{person}{Shyong~K. Lam} {and} \bibinfo{person}{John
  Riedl}.} \bibinfo{year}{2004}\natexlab{}.
\newblock \showarticletitle{Shilling recommender systems for fun and profit}.
  In \bibinfo{booktitle}{\emph{Proceedings of the 13th international conference
  on World Wide Web, {WWW} 2004, New York, NY, USA, May 17-20, 2004}},
  \bibfield{editor}{\bibinfo{person}{Stuart~I. Feldman}, \bibinfo{person}{Mike
  Uretsky}, \bibinfo{person}{Marc Najork}, {and} \bibinfo{person}{Craig~E.
  Wills}} (Eds.). \bibinfo{publisher}{{ACM}}, \bibinfo{pages}{393--402}.
\newblock
\showISBNx{1-58113-844-X}
\urldef\tempurl%
\url{https://doi.org/10.1145/988672.988726}
\showDOI{\tempurl}


\bibitem[\protect\citeauthoryear{Linden, Smith, and York}{Linden
  et~al\mbox{.}}{2003}]%
        {DBLP:journals/dsonline/LindenSY03}
\bibfield{author}{\bibinfo{person}{Greg Linden}, \bibinfo{person}{Brent Smith},
  {and} \bibinfo{person}{Jeremy York}.} \bibinfo{year}{2003}\natexlab{}.
\newblock \showarticletitle{Industry Report: Amazon.com Recommendations:
  Item-to-Item Collaborative Filtering}.
\newblock \bibinfo{journal}{\emph{{IEEE} Distributed Systems Online}}
  \bibinfo{volume}{4}, \bibinfo{number}{1} (\bibinfo{year}{2003}).
\newblock


\bibitem[\protect\citeauthoryear{Mehta and Nejdl}{Mehta and Nejdl}{2008}]%
        {DBLP:conf/sigir/MehtaN08}
\bibfield{author}{\bibinfo{person}{Bhaskar Mehta} {and}
  \bibinfo{person}{Wolfgang Nejdl}.} \bibinfo{year}{2008}\natexlab{}.
\newblock \showarticletitle{Attack resistant collaborative filtering}. In
  \bibinfo{booktitle}{\emph{Proceedings of the 31st Annual International {ACM}
  {SIGIR} Conference on Research and Development in Information Retrieval,
  {SIGIR} 2008, Singapore, July 20-24, 2008}},
  \bibfield{editor}{\bibinfo{person}{Sung{-}Hyon Myaeng},
  \bibinfo{person}{Douglas~W. Oard}, \bibinfo{person}{Fabrizio Sebastiani},
  \bibinfo{person}{Tat{-}Seng Chua}, {and} \bibinfo{person}{Mun{-}Kew Leong}}
  (Eds.). \bibinfo{publisher}{{ACM}}, \bibinfo{pages}{75--82}.
\newblock
\showISBNx{978-1-60558-164-4}
\urldef\tempurl%
\url{https://doi.org/10.1145/1390334.1390350}
\showDOI{\tempurl}


\bibitem[\protect\citeauthoryear{Mehta and Nejdl}{Mehta and Nejdl}{2009}]%
        {DBLP:journals/umuai/MehtaN09}
\bibfield{author}{\bibinfo{person}{Bhaskar Mehta} {and}
  \bibinfo{person}{Wolfgang Nejdl}.} \bibinfo{year}{2009}\natexlab{}.
\newblock \showarticletitle{Unsupervised strategies for shilling detection and
  robust collaborative filtering}.
\newblock \bibinfo{journal}{\emph{User Model. User-Adapt. Interact.}}
  \bibinfo{volume}{19}, \bibinfo{number}{1-2} (\bibinfo{year}{2009}),
  \bibinfo{pages}{65--97}.
\newblock
\urldef\tempurl%
\url{https://doi.org/10.1007/s11257-008-9050-4}
\showDOI{\tempurl}


\bibitem[\protect\citeauthoryear{Mobasher, Burke, Bhaumik, and
  Williams}{Mobasher et~al\mbox{.}}{2005}]%
        {mobasher2005effective}
\bibfield{author}{\bibinfo{person}{Bamshad Mobasher}, \bibinfo{person}{Robin
  Burke}, \bibinfo{person}{Runa Bhaumik}, {and} \bibinfo{person}{Chad
  Williams}.} \bibinfo{year}{2005}\natexlab{}.
\newblock \showarticletitle{Effective attack models for shilling item-based
  collaborative filtering systems}. Citeseer.
\newblock


\bibitem[\protect\citeauthoryear{Mobasher, Burke, and Sandvig}{Mobasher
  et~al\mbox{.}}{2006}]%
        {mobasher2006model}
\bibfield{author}{\bibinfo{person}{Bamshad Mobasher}, \bibinfo{person}{Robin
  Burke}, {and} \bibinfo{person}{Jeff~J Sandvig}.}
  \bibinfo{year}{2006}\natexlab{}.
\newblock \showarticletitle{Model-based collaborative filtering as a defense
  against profile injection attacks}. In \bibinfo{booktitle}{\emph{AAAI}},
  Vol.~\bibinfo{volume}{6}. \bibinfo{pages}{1388}.
\newblock


\bibitem[\protect\citeauthoryear{Mobasher, Burke, Bhaumik, and
  Sandvig}{Mobasher et~al\mbox{.}}{2007a}]%
        {DBLP:journals/expert/MobasherBBS07}
\bibfield{author}{\bibinfo{person}{Bamshad Mobasher}, \bibinfo{person}{Robin~D.
  Burke}, \bibinfo{person}{Runa Bhaumik}, {and} \bibinfo{person}{Jeff~J.
  Sandvig}.} \bibinfo{year}{2007}\natexlab{a}.
\newblock \showarticletitle{Attacks and Remedies in Collaborative
  Recommendation}.
\newblock \bibinfo{journal}{\emph{{IEEE} Intelligent Systems}}
  \bibinfo{volume}{22}, \bibinfo{number}{3} (\bibinfo{year}{2007}),
  \bibinfo{pages}{56--63}.
\newblock
\urldef\tempurl%
\url{https://doi.org/10.1109/MIS.2007.45}
\showDOI{\tempurl}


\bibitem[\protect\citeauthoryear{Mobasher, Burke, Bhaumik, and
  Williams}{Mobasher et~al\mbox{.}}{2007b}]%
        {DBLP:journals/toit/MobasherBBW07}
\bibfield{author}{\bibinfo{person}{Bamshad Mobasher}, \bibinfo{person}{Robin~D.
  Burke}, \bibinfo{person}{Runa Bhaumik}, {and} \bibinfo{person}{Chad
  Williams}.} \bibinfo{year}{2007}\natexlab{b}.
\newblock \showarticletitle{Toward trustworthy recommender systems: An analysis
  of attack models and algorithm robustness}.
\newblock \bibinfo{journal}{\emph{{ACM} Trans. Internet Techn.}}
  \bibinfo{volume}{7}, \bibinfo{number}{4} (\bibinfo{year}{2007}),
  \bibinfo{pages}{23}.
\newblock
\urldef\tempurl%
\url{https://doi.org/10.1145/1278366.1278372}
\showDOI{\tempurl}


\bibitem[\protect\citeauthoryear{Ning and Karypis}{Ning and Karypis}{2011}]%
        {DBLP:conf/icdm/NingK11}
\bibfield{author}{\bibinfo{person}{Xia Ning} {and} \bibinfo{person}{George
  Karypis}.} \bibinfo{year}{2011}\natexlab{}.
\newblock \showarticletitle{{SLIM:} Sparse Linear Methods for Top-N Recommender
  Systems}. In \bibinfo{booktitle}{\emph{11th {IEEE} International Conference
  on Data Mining, {ICDM} 2011, Vancouver, BC, Canada, December 11-14, 2011}},
  \bibfield{editor}{\bibinfo{person}{Diane~J. Cook}, \bibinfo{person}{Jian
  Pei}, \bibinfo{person}{Wei Wang}, \bibinfo{person}{Osmar~R. Za{\"{\i}}ane},
  {and} \bibinfo{person}{Xindong Wu}} (Eds.). \bibinfo{publisher}{{IEEE}
  Computer Society}, \bibinfo{pages}{497--506}.
\newblock
\showISBNx{978-0-7695-4408-3}
\urldef\tempurl%
\url{https://doi.org/10.1109/ICDM.2011.134}
\showDOI{\tempurl}


\bibitem[\protect\citeauthoryear{O'Mahony, Hurley, Kushmerick, and
  Silvestre}{O'Mahony et~al\mbox{.}}{2004}]%
        {DBLP:journals/toit/OMahonyHKS04}
\bibfield{author}{\bibinfo{person}{Michael~P. O'Mahony},
  \bibinfo{person}{Neil~J. Hurley}, \bibinfo{person}{Nicholas Kushmerick},
  {and} \bibinfo{person}{Guenole C.~M. Silvestre}.}
  \bibinfo{year}{2004}\natexlab{}.
\newblock \showarticletitle{Collaborative recommendation: {A} robustness
  analysis}.
\newblock \bibinfo{journal}{\emph{{ACM} Trans. Internet Techn.}}
  \bibinfo{volume}{4}, \bibinfo{number}{4} (\bibinfo{year}{2004}),
  \bibinfo{pages}{344--377}.
\newblock
\urldef\tempurl%
\url{https://doi.org/10.1145/1031114.1031116}
\showDOI{\tempurl}


\bibitem[\protect\citeauthoryear{O'Mahony, Hurley, and Silvestre}{O'Mahony
  et~al\mbox{.}}{2005}]%
        {DBLP:conf/aaai/OMahonyHS05}
\bibfield{author}{\bibinfo{person}{Michael~P. O'Mahony},
  \bibinfo{person}{Neil~J. Hurley}, {and} \bibinfo{person}{Guenole C.~M.
  Silvestre}.} \bibinfo{year}{2005}\natexlab{}.
\newblock \showarticletitle{Recommender Systems: Attack Types and Strategies}.
  In \bibinfo{booktitle}{\emph{Proceedings, The Twentieth National Conference
  on Artificial Intelligence and the Seventeenth Innovative Applications of
  Artificial Intelligence Conference, July 9-13, 2005, Pittsburgh,
  Pennsylvania, {USA}}}, \bibfield{editor}{\bibinfo{person}{Manuela~M. Veloso}
  {and} \bibinfo{person}{Subbarao Kambhampati}} (Eds.).
  \bibinfo{publisher}{{AAAI} Press / The {MIT} Press},
  \bibinfo{pages}{334--339}.
\newblock
\showISBNx{1-57735-236-X}
\urldef\tempurl%
\url{http://www.aaai.org/Library/AAAI/2005/aaai05-053.php}
\showURL{%
\tempurl}


\bibitem[\protect\citeauthoryear{Rendle, Freudenthaler, Gantner, and
  Schmidt{-}Thieme}{Rendle et~al\mbox{.}}{2009}]%
        {DBLP:conf/uai/RendleFGS09}
\bibfield{author}{\bibinfo{person}{Steffen Rendle}, \bibinfo{person}{Christoph
  Freudenthaler}, \bibinfo{person}{Zeno Gantner}, {and} \bibinfo{person}{Lars
  Schmidt{-}Thieme}.} \bibinfo{year}{2009}\natexlab{}.
\newblock \showarticletitle{{BPR:} Bayesian Personalized Ranking from Implicit
  Feedback}. In \bibinfo{booktitle}{\emph{{UAI} 2009, Proceedings of the
  Twenty-Fifth Conference on Uncertainty in Artificial Intelligence, Montreal,
  QC, Canada, June 18-21, 2009}}, \bibfield{editor}{\bibinfo{person}{Jeff~A.
  Bilmes} {and} \bibinfo{person}{Andrew~Y. Ng}} (Eds.).
  \bibinfo{publisher}{{AUAI} Press}, \bibinfo{pages}{452--461}.
\newblock
\urldef\tempurl%
\url{https://dslpitt.org/uai/displayArticleDetails.jsp?mmnu=1\&smnu=2\&article\_id=1630\&proceeding\_id=25}
\showURL{%
\tempurl}


\bibitem[\protect\citeauthoryear{Sarwar, Karypis, Konstan, and Riedl}{Sarwar
  et~al\mbox{.}}{2001}]%
        {DBLP:conf/www/SarwarKKR01}
\bibfield{author}{\bibinfo{person}{Badrul~Munir Sarwar},
  \bibinfo{person}{George Karypis}, \bibinfo{person}{Joseph~A. Konstan}, {and}
  \bibinfo{person}{John Riedl}.} \bibinfo{year}{2001}\natexlab{}.
\newblock \showarticletitle{Item-based collaborative filtering recommendation
  algorithms}. In \bibinfo{booktitle}{\emph{Proceedings of the Tenth
  International World Wide Web Conference, {WWW} 10, Hong Kong, China, May 1-5,
  2001}}, \bibfield{editor}{\bibinfo{person}{Vincent~Y. Shen},
  \bibinfo{person}{Nobuo Saito}, \bibinfo{person}{Michael~R. Lyu}, {and}
  \bibinfo{person}{Mary~Ellen Zurko}} (Eds.). \bibinfo{publisher}{{ACM}},
  \bibinfo{pages}{285--295}.
\newblock
\showISBNx{1-58113-348-0}
\urldef\tempurl%
\url{https://doi.org/10.1145/371920.372071}
\showDOI{\tempurl}


\bibitem[\protect\citeauthoryear{Shi, Larson, and Hanjalic}{Shi
  et~al\mbox{.}}{2014}]%
        {shi2014collaborative}
\bibfield{author}{\bibinfo{person}{Yue Shi}, \bibinfo{person}{Martha Larson},
  {and} \bibinfo{person}{Alan Hanjalic}.} \bibinfo{year}{2014}\natexlab{}.
\newblock \showarticletitle{Collaborative filtering beyond the user-item
  matrix: A survey of the state of the art and future challenges}.
\newblock \bibinfo{journal}{\emph{ACM Computing Surveys (CSUR)}}
  \bibinfo{volume}{47}, \bibinfo{number}{1} (\bibinfo{year}{2014}),
  \bibinfo{pages}{3}.
\newblock


\bibitem[\protect\citeauthoryear{Yang and Cai}{Yang and Cai}{2017}]%
        {DBLP:journals/jiis/YangC17}
\bibfield{author}{\bibinfo{person}{Zhihai Yang} {and} \bibinfo{person}{Zhongmin
  Cai}.} \bibinfo{year}{2017}\natexlab{}.
\newblock \showarticletitle{Detecting abnormal profiles in collaborative
  filtering recommender systems}.
\newblock \bibinfo{journal}{\emph{J. Intell. Inf. Syst.}} \bibinfo{volume}{48},
  \bibinfo{number}{3} (\bibinfo{year}{2017}), \bibinfo{pages}{499--518}.
\newblock
\urldef\tempurl%
\url{https://doi.org/10.1007/s10844-016-0424-5}
\showDOI{\tempurl}


\bibitem[\protect\citeauthoryear{Zhang, Tang, Zhang, and Xue}{Zhang
  et~al\mbox{.}}{2014}]%
        {zhang2014addressing}
\bibfield{author}{\bibinfo{person}{Mi Zhang}, \bibinfo{person}{Jie Tang},
  \bibinfo{person}{Xuchen Zhang}, {and} \bibinfo{person}{Xiangyang Xue}.}
  \bibinfo{year}{2014}\natexlab{}.
\newblock \showarticletitle{Addressing cold start in recommender systems: A
  semi-supervised co-training algorithm}. In
  \bibinfo{booktitle}{\emph{Proceedings of the 37th international ACM SIGIR
  conference on Research \& development in information retrieval}}. ACM,
  \bibinfo{pages}{73--82}.
\newblock


\end{thebibliography}

\end{document}